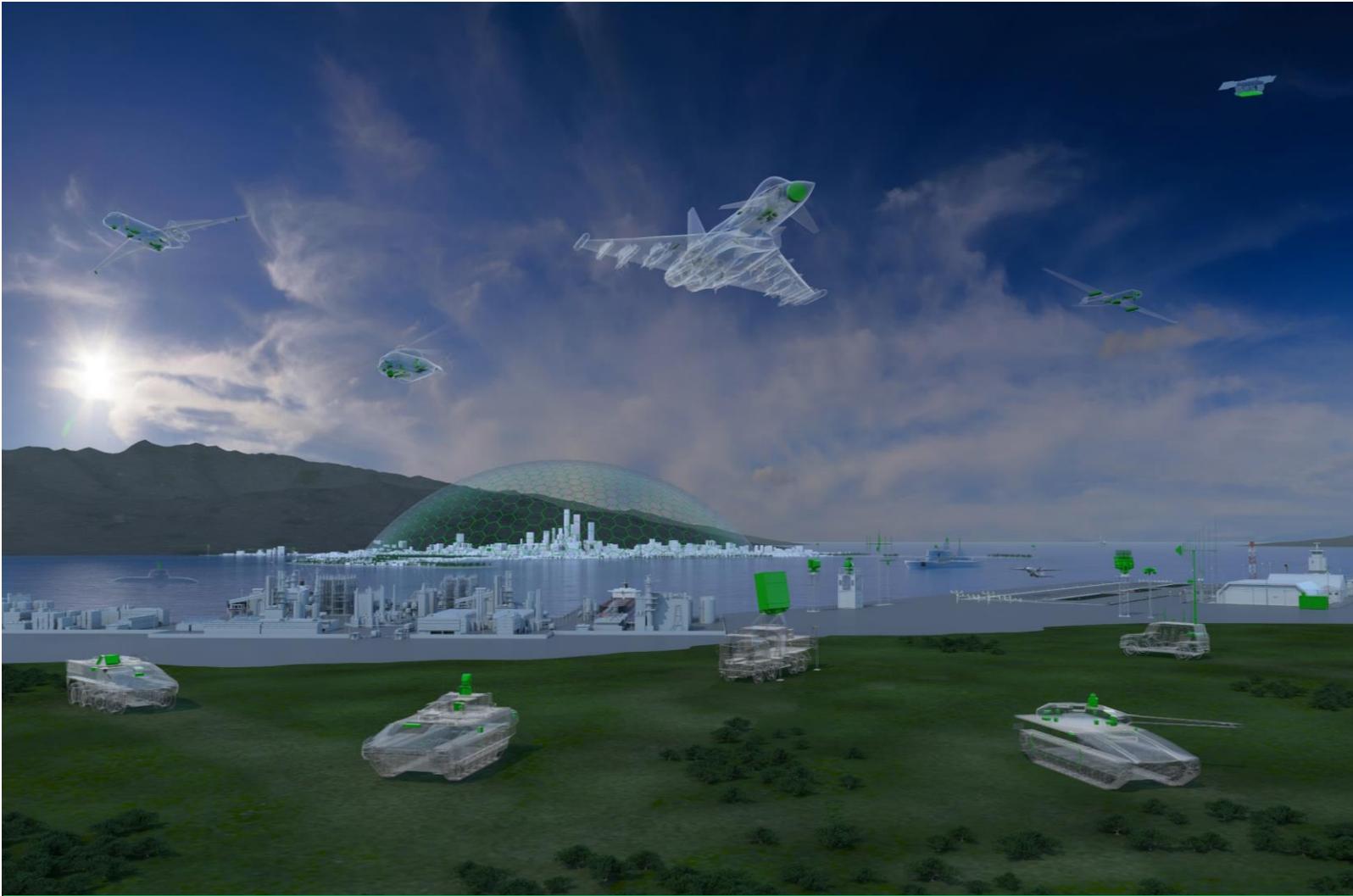

White Paper

# Ethical Considerations for the Military Use of Artificial Intelligence in Visual Reconnaissance

31 January 2025

Fraunhofer IOSB

HENSOLDT Detect and Protect.

# Table of Contents

# List of Figures

## Authors

| | |
|---|---|
| Dr. Mathias Anneken | HENSOLDT Sensors, Ulm, Germany |
| Dr. Nadia Burkart | Fraunhofer IOSB, Karlsruhe, Germany |
| Fabian Jeschke | HENSOLDT Optronics, Oberkochen, Germany |
| Dr. Achim Kuwertz-Wolf | Fraunhofer IOSB, Karlsruhe, Germany |
| Dr. Almuth Müller | Fraunhofer IOSB, Karlsruhe, Germany |
| Dr. Arne Schumann | Fraunhofer IOSB, Karlsruhe, Germany |
| Dr. Michael Teutsch | HENSOLDT Sensors, Ulm, Germany |

All authors contributed equally to this work.

# 1 Introduction

Artificial Intelligence (AI) and related technologies in machine and deep learning are making their way into all of our lives: intelligent dialog systems and smartphones powered by AI simplify everyday life, self-driving vehicles with AI-based signal processing and control increase the level of automation, and AI is considered as one of the building blocks of the fourth industrial revolution. However, while AI in general becomes more powerful and widely used, the impact on humans and their well-being must be considered. It can thus no longer be sufficient to just technically enable AI to solve a given task. Its use has to follow guidelines for human values and ethical principles. This implication not only holds for the general use of AI, but it is especially relevant for military applications. In this white paper, we analyse and discuss how the responsible use of AI in military scenarios can be achieved with compliance to ethical and legal guidelines. In Section 2, we first review existing literature that broadly discusses ethical guidelines for responsible AI in the civil and the military domain. Based on these findings, we derive a consolidated set of ethical principles that reflects our opinion towards an ethically compliant integration of AI-powered assistance functions in military applications.

Related work especially in the military domain currently focuses on rather theoretical considerations of this topic [4, 23]. In Sections 3, 4 and 5, however, we describe three specific use cases, where AI-powered assistance functions are utilized in military applications. Each specific scenario is described along with a technical point of view that shows the feasibility of the AI-based functions with current state-of-the-art technology. Furthermore, our derived ethical principles are mapped to specific observations and recommendations of the AI-powered system as well as interactions with the user, like human-in-the-loop and human-on-the-loop principles. In Section 6, we conclude that a careful design and development of AI-powered assistance systems together with suitable integrated methods of eXplainable Artificial Intelligence (XAI) can support the human user substantially and it can be well-aligned with ethical principles. Especially in military applications, the human-in-the-loop shall still be the final and sole decision-maker.

# 2 Principles of Ethical AI

This section begins with a review of ethical principles for both the civilian and military domains. Section 2.1, presents a short review on ethical principles for AI in the civil domain. Subsequently, Section 2.2 reviews ethical principles for AI in the military domain with a focus on decision support systems.

Building upon those reviewed principles, consolidated ethical principles are introduced in Section 2.3, which will subsequently serve to evaluate the presented use cases (Sections 3, 4, and 5) with respect to the alignment and implementation of ethical AI.

## 2.1 Review of Considerations for Responsible AI for the General Use

In the literature, several guidelines exist such as Fairness, Accountability, Transparency, and Ethics (FATE) in AI[1][2][3]. A meta-analysis on guidelines for ethical AI is given, e.g., in [10]. This meta-analysis includes the viewpoints of companies, academia and governments from around the globe. Jobin et al. [10] found the following four main topics in the different guidelines:

**Transparency:** describes the need for AI systems to be transparent in their decision-making processes and the data they use. Transparency also relates to terms such as explainability, understandability, interpretability, or disclosure.

**Justice & fairness:** this refers to the idea that AI systems should be designed to treat all individuals fairly and without discrimination. They also refer to terms such as consistency, inclusion, equality, (non-)bias, (non-)discrimination, diversity, or plurality.

**Non-maleficence:** states that AI systems should not cause harm to individuals or society. This includes both physical harm and harm to privacy and civil liberties. Non-maleficence also refers to terms such as security and safety, protection, precaution, or integrity (bodily or mental).

**Responsibility:** stands for the obligation of those involved in the development, deployment, and use of AI systems to ensure that they are used ethically and responsibly. This includes taking steps to prevent harm, as well as being accountable for the consequences of AI systems. Responsibility also refers to terms such as accountability, liability, or acting with integrity.

While these topics align with the FATE principles, non-maleficence is explicitly stated. Another topic indicated by the meta-analysis, which is generally less emphasized, is the concept of freedom & autonomy: this concept refers to the idea that individuals should have the freedom to make their own choices and have control over the data and decisions that AI systems make about them. This involves ensuring that individuals are not subject to automated decision-making processes that restrict their freedom or autonomy without just cause. Both the principles of non-maleficence and freedom & autonomy present challenging ethical considerations in the use of AI in military scenarios, requiring careful consideration and a delicate balance.

The aforementioned considerations are about general aspects of ethical AI. Therefore, they serve as the foundation for the specific examination in our later presented and discussed use cases. However, there are additional crucial aspects to consider, such as responsibility and accountability, specifically in the military context. Hence, we will delve deeper into the existing guidelines for the utilization of AI in the military domain before proceeding further. At this point, we want to emphasize that throughout human history, extensive considerations have been given to military operations and ethics. One notable framework in this regard is the Just War Theory, which is a doctrine or tradition of military ethics. The Just War Theory establishes a set of criteria that must be met for a war to be deemed morally justifiable.

---

[1] https://www.microsoft.com/en-us/research/theme/fate/  2http://www.cycat.io/introduction-to-fate-in-ai/
[2] http://cycat.io/introduction-to-fate-in-ai/
[3] https://assets.kpmg.com/content/dam/kpmg/au/pdf/2019/artificial-intelligence-perspectives-on-fate-in-ai.pdf

## 2.2 Review of Ethical Considerations for AI in Military Use

The relevance of harnessing AI in defence is increasing more and more. The dedicated ethical discussion, however, is still coming short in such examinations to analyse specific applications like in the fielding for armed forces [2]. The analysis so far is mainly based on the civilian approach in defining ethical AI and giving guidelines for these use cases. An exception is a position paper [7], which addresses current issues of trustworthy AI and identifies the need to further define and research concepts for the responsible use of AI in autonomous weapon systems. As mentioned before, AI systems for defence applications are subject to the same principles as civilian ones with a special focus on certain defence-related aspects. In practice, e.g., the United States Department of Defence (DoD) officially adopted ethical principles for AI in 2020 within a "Responsible Artificial Intelligence Strategy and Implementation Pathway" [4]. A similar approach is provided by the North Atlantic Treaty Organization (NATO). A further analysis of these principles, here based on the Just War Theory, is given in [23]. The following sections will give an overview of those highly relevant works.

**AI Ethical Principles by the U.S. Department of Defence (DoD)**

The five ethical principles highlighted by the DoD are based on AI being Responsible, Equitable, Traceable, Reliable, and Governable. In the following paragraphs, these principles are further defined.

**Responsible:** the responsible use of AI systems requires human beings to exercise appropriate levels of judgment and accountability throughout the development, deployment, use, and outcomes of these systems. This means that individuals must take responsibility for ensuring that the AI systems do not harm people, damage property, or violate ethical, legal, or moral principles. This requires not only technical expertise but also a deep understanding of the broader social, cultural, and ethical implications of AI.

**Equitable:** in developing and deploying AI systems, it is important to ensure that these systems do not cause unintended harm to individuals or groups. This requires taking deliberate steps to avoid unintended bias and discrimination in the development and deployment of combat or non-combat AI systems. It means ensuring that these systems are developed and trained on diverse datasets and that they do not perpetuate existing biases or prejudices. This also requires ongoing monitoring and evaluation of these systems to ensure that they do not cause harm to people.

**Traceable:** the AI engineering discipline must be sufficiently advanced such that technical experts possess an appropriate understanding of the technology, development processes, and operational methods of AI systems. This includes transparent and auditable methodologies, data sources, and design procedures and documentation. The traceability of AI systems is crucial to understand how they operate, to identify potential risks and vulnerabilities, and to ensure that they comply with ethical and legal requirements.

**Reliable:** AI systems must be reliable, safe, secure, and robust throughout their entire lifecycle within their defined domain of use. This means that these systems must be designed and tested to perform their intended function under expected and unexpected conditions. It requires rigorous testing and evaluation to ensure that these systems are safe and secure, and to identify and mitigate potential risks or vulnerabilities. The reliability of AI systems is critical to ensure that they perform their intended functions accurately and effectively, without causing harm to individuals or property.

**Governable:** AI systems must be designed and engineered to fulfil their intended function while possessing the ability to detect and avoid unintended harm or disruption. This requires a system that can be governed effectively, both by humans and by automated systems. This includes the ability to (1) monitor and control the behaviour of deployed AI systems, and to (2) deactivate or disengage these systems if they demonstrate unintended escalatory or other unwanted behaviour. The governability of AI systems is essential to ensure that they operate within ethical and legal boundaries and do not cause harm to individuals or groups.

**AI Ethical Principles by the North Atlantic Treaty Organization (NATO)**

Relatively similar to the principles of the DoD is the strategy of the NATO for incorporating AI into defence and security operations, underscoring the profound impact of AI on global defence [15]. The strategy seeks to set a responsible example in the development and application of AI for defence and security, with four primary objectives: mainstreaming AI adoption, safeguarding AI technologies, countering malicious AI use, and promoting NATO's Principles of Responsible Use. The principles encompass aspects such as lawfulness, responsibility, accountability, explainability, traceability, reliability, governability, and bias mitigation. The strategy also addresses the secure and responsible deployment of AI, interference minimization, and the potential repercussions on critical infrastructure.

**AI Ethical Principles by the U.K. Defence Science and Technology Lab (DSTL)**

The paper by Taddeo [23] is notable among the literature on ethical AI principles, particularly in the military context, as it states the need to align ethical principles with the intended purpose of military action. Rather than adopting a one-size-fits-all approach to ethical considerations, the specificities of military operations need to be taken into account. Those aspects can be expressed by the three core categories of military action as defined by defence institutions: support and sustainment, non-kinetic adversarial actions, and kinetic adversarial engagements. The use of AI to sustain and support refers to instances where AI is utilized for administrative functions, resource distribution, and enhancing the security of communication and infrastructure systems that support national defence. Adversarial and non-kinetic uses of AI encompass a range of applications, from countering cyber threats to carrying out offensive cyber operations with non-lethal goals. Adversarial and kinetic uses involve integrating AI systems into combat operations, such as using AI to assist in identifying targets or using lethal autonomous weapon systems (LAWS). By doing this, Taddeo has set the stage for a more nuanced analysis of military AI ethics. However, the paper intentionally limits its scope to the first two categories and does not delve into the third category—adversarial and kinetic uses of AI—which is deferred for a detailed examination in subsequent papers [1, 22]. In formulating their principles, [23] considered the principles proposed by the US Defence Innovation Board [5] in 2020, which influenced the development of the US Department of Defence's recently updated principles for military AI use, particularly in non-kinetic operations.

Thus, we also consider these principles with respect to the decision support aspect of our three use cases. This analysis and consideration of those principles is crucial in ensuring that the use cases adheres to the highest ethical standards and are consistent with the latest academic discourse on the subject. A summary of the principles by Taddeo [23]:

**Justified and overridable uses:** in defence, AI systems must serve a legitimate military objective and comply with international humanitarian law. They require human oversight and control to ensure ethical principles are upheld, human rights are not violated, and international peace and security are not undermined.

**Just and transparent systems and processes:** to maintain ethical standards, transparency and accountability are crucial for AI in defence systems. This can be achieved by implementing ethical auditing processes, deploying systems in accordance with the principles of the Just War theory, and maintaining traceability throughout the development and deployment phases. These measures ensure that ethical breaches can be identified, and responsibility attributed, promoting ethical scrutiny and soundness of outcomes.

**Human moral responsibility:** humans are ultimately responsible for the outcomes of AI systems deployed in defence, as AI cannot be held morally accountable. To address this, a linear or radial approach can be taken to identify and address unwanted consequences and foster a self-improving dynamic in the network of agents involved in AI development and deployment for defence.

**Meaningful human control:** the principle of meaningful human control ensures that humans have the ability to intervene or deactivate AI systems promptly and exercise control in a situation-dependent way. This principle can be implemented in various degrees, depending on technical, legal, and ethical training.

However, it requires rigorous risk assessments to identify unintended consequences and negative impacts, and protocols for attributing responsibilities and ensuring transparency.

**Reliable AI systems:** to ensure reliable AI systems for defence, meaningful monitoring is crucial. As AI lacks transparency and robustness, monitoring throughout deployment is necessary. This can include in-house development, adversarial training, and monitoring in the wild with baseline models. Monitoring should span from design to deployment.

## 2.3 Consolidation of Ethical Principles for Decision-supporting AI in Military Use

In the following, we present our ethical principles, which we have derived through the analysis of the principles summarized in Section 2.1 and 2.2. Our principles reflect our understanding of the underlying semantic concepts behind the terms and have been formulated with consideration of our three use cases.

**Traceability**

We can find the principle of traceability in just and transparent systems and processes in [23] and in being traceable in [4]. One key aspect of traceability is transparency: a human user has to be able to not only understand a result presented by an AI-based system, but also the process how the result was derived. To this end, methods for XAI can be employed, but only to a certain level. Since AI-based decision support systems are often multi-layered (cf. hybrid AI systems), transparency has to be provided not only at each individual layer such as artificial neural networks or rule-based AI, but also in their combination. This is highly challenging, however, explanations of results and checking their plausibility can be important steps for transparent decision support systems. Furthermore, traceability also considers and encourages to foster the general understanding of the AI system by humans, not only during its deployment, but also during its development already, e.g. with concepts such as security-by-design or safety-by-design. Thus, traceability not only involves system operators, but also designers and developers of AI systems. Finally, traceability for us also relates to accountability (e.g., who is responsible in case of erroneous decisions) and therefore encompasses aspects of monitoring, auditing, and logging.

**Proportionality**

Proportionality is of a cross-cutting nature regarding the above-mentioned principles. As a vital part of the Just War theory, it relates to or is partially addressed in the principles (1) justified and overridable uses, (2) just and transparent systems and processes, (3) human moral responsibility, and (4) meaningful human control [23]. It concerns the results of an AI-based decision support system being aligned with humanitarian law, a legitimate military objective, and the corresponding Rules of Engagement (RoE). More specifically, results provided by an AI system must also be considered within the current situational context, weighing detected threats versus the risks of proposed countermeasures. Usually, this will involve a form of human control. Proportionality closely relates to necessity and discrimination within the Just War theory. It also addresses the ethical principles of general AI of non-maleficence and freedom & autonomy from a military perspective, as previously identified as important for the military use of AI. We thus consider proportionality as a crucial ethical principle for our use cases, since it can be a fundamental part in a nation's military doctrine or in international missions led for instance by the United Nations (UN). In a similar fashion, it is considered as important for all adversarial use cases (i.e. kinetic and non-kinetic ones) as discussed in [23].

**Governability**

The often-used term governance does not comprehensively represent our focus: since enabling governance should be the intention of an ethical principle, we consider governability as the ability to enforce governance over the AI system. Furthermore, the principle of governability is presented in being governable [4] and related to or addressed in justified and overridable uses as well as meaningful human control in [23]. Here, an employed AI-based system must be controllable in the sense that a human user is able to override an AI-based decision at any point or completely deactivate the respective system. For our use cases, the human-in-the-loop approach is thus an important aspect of governability. As a prerequisite, the system must be traceable for a human user to understand the AI results and their consequences and act accordingly. Consequently, the responsibility lies with the human user.

**Responsibility**

We consider the principle of responsibility as presented in human moral responsibility [23] and being responsible [4]. For a military use of AI in decision support systems, the human user as decision maker is ultimately responsible. The human-in-the-loop approach is therefore paramount: AI methods support a human user and are thus part of an AI-based assistance system. Users in turn have to understand the AI in the best possible ways (cf. the principles traceability and reliability). This principle, however, is not only relevant for human end users of an AI-based system: it also encompasses the responsibility of machine learning engineers for developing the AI methods.

**Reliability**

We consider the principle of reliability as presented in reliable AI systems [23] and being reliable [4]. This contains the robustness of AI methods in defined use cases regarding e.g. high detection probabilities for objects and adequate false alarm rates, as well as feedback to human users in cases, where the AI cannot reach a reliable result such as low confidence results or ambiguity. In addition, the principle of being equitable [4] has a part to play here: if there is an unattended bias in the training data used for machine learning, this may lead to erroneous results and thus less reliable AI methods.

All principles, including our own aggregation and interpretation, are valuable and warrant careful semantic definition. However, the focus has to shift to how these principles are implemented and manifested in action. It raises the questions of how AI is utilized in military systems and how does responsible AI present itself to the user in such systems.

## 3 Use Case 1 - Decision Support for Maritime Surveillance

### 3.1 Scenario

The first scenario takes place in a maritime setting. A submarine is on a surveillance mission. The submarine scans the surrounding area with an optronic mast that provides an automated 360° scan. Threats from enemy military vessels are unlikely but possible. An officer analyses the image and is supported by an AI system. The AI system contains an automatic object detection and classification function together with a rule set for maritime threat scenarios, like attacks or collisions.

In Figure 1, a schematic Graphical User Interface (GUI) for the surveillance software is shown.

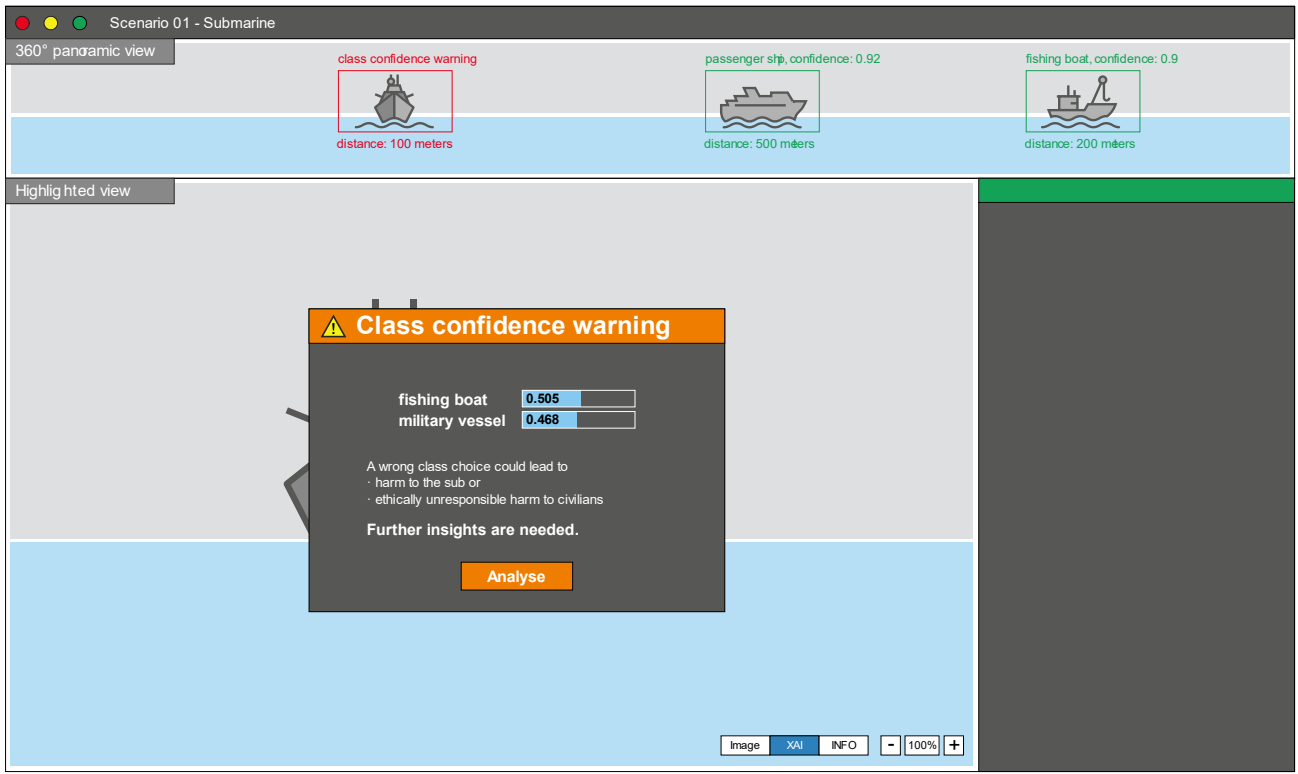

Figure 1  Schematic GUI for a submarine surveillance AI software. The screen is divided into three parts: a panoramic view on top showing the 360° periscope image, a highlight view in the middle, and a side panel on the right for further information. The main screen is overlayed with a class confidence warning, indicating to the user that the object classification AI needs support from the user.

In the upper part, the officer sees the full image from the periscope as a panoramic view. An object classification method analyses the image and frames each identified man-made object with a rectangle bounding box. Around each bounding box is the distance of the object from the submarine and its classification together with the confidence of this classification. Three vessels are detected, in line with the earlier sonar scan, and two of them are labelled as civilian vessels, therefore posing no threat to the submarine. The object classification is unable to classify the third vessel as either civilian or military. This is shown to the officer with a red bounding box around that object. Simultaneously, a class confidence warning pops up on the screen to alert the officer. The classification method shows both labels' confidence as graphics so that the officer can understand the problem better. Alongside the label predictions, there's also information about the severance of the choice for either of the class: a wrong class choice could lead to either harm to the submarine or ethically unresponsive harm to civilians. The AI system method points out that there's more information needed to make an insightful decision by the officer.

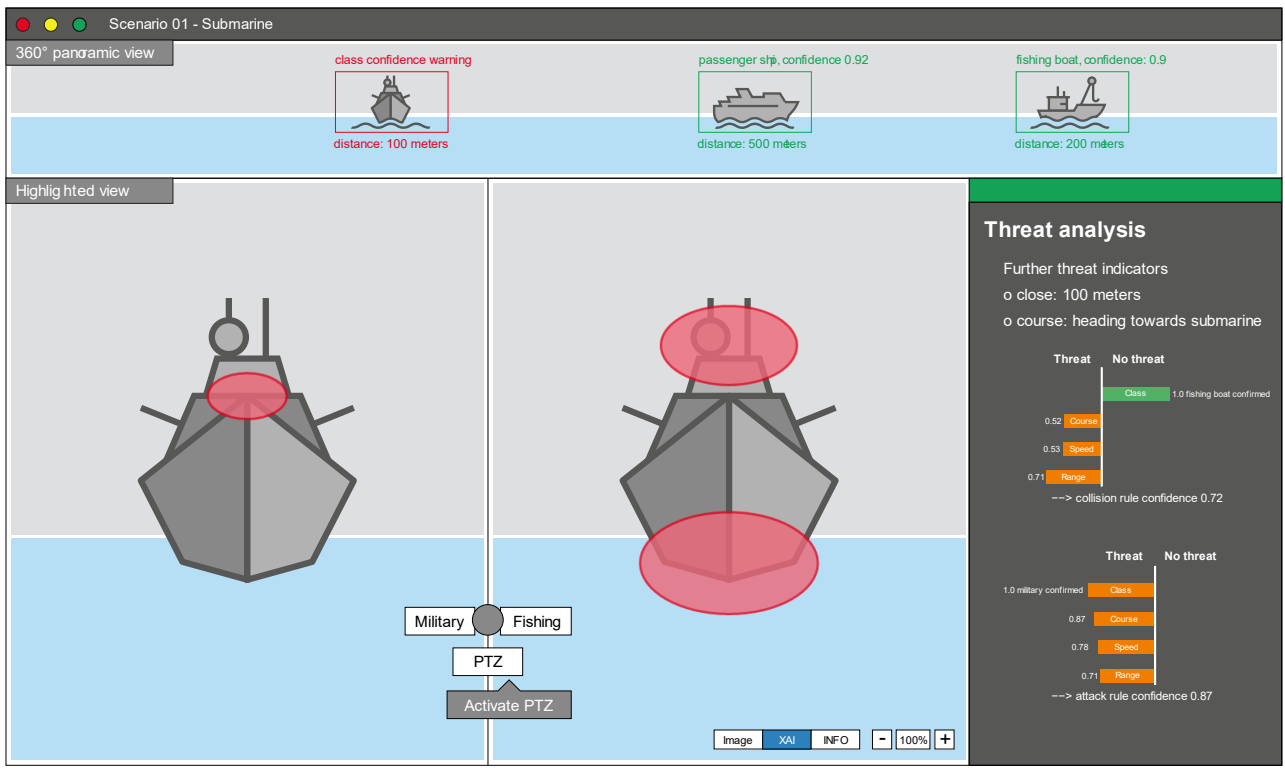

Figure 2  Schematic GUI for a submarine surveillance AI software. The screen is divided into three parts: a panoramic view on top showing the 360° periscope image, a highlight view in the middle showing an XAI view with heatmap overlays for each predicted label, and a side panel on the right for further threat analysis information.

The officer confirms that he has seen the warning by clicking on the analyse button. Internally the software is logging the officer's action for later revisions. After clicking on the analyse button, an XAI view opens (see Figure 2). Here, the officer can see the important parts of the boat that indicate the labels as a heatmap overlay. The panel on the right side gives some further information about the threat analysis. Here, the officer can see the important indicators for a potential navigation threat. Those are the range of the vessel and its course and speed. Below that two simulations are shown for each of the possible labels. If the officer is sure, it is a fishing boat, then the indicators do not lead to a threat, but a collision rule applies, because the boat is still heading towards the submarine, and is quite close yet. Conversely, if the officer is sure, it is a military vessel, then the indicators lead to an attack prediction.

The officer sees that the front part of the boat was important for the military label. The officer uses digital zoom to narrow in on the suspicious vessel. With this close-up view, the officer can identify the important part at the front of the vessel as a harpoon used in whale hunting. Therefore, the officer can identify the boat as a fishing boat. To avoid collision of the submarine with the fishing boat, the officer orders diving to a safe depth.

## 3.2 Technical Point of View

In order to emphasize the level of technical realism of this scenario, we outline a possible implementation through state-of-the-art AI and XAI methods. Exemplary results are demonstrated on a specifically chosen data sample of a whaling ship shown in Figure 3.

During the periscope scan, vessels of all types are first detected using a modern object detection approach based on deep learning and deep neural networks. A well-established approach to employ for this is the Fully Convolutional One-Stage (FCOS) detection model [24, 21]. Several publicly available datasets, such as the Singapore Maritime Dataset [17, 14], provide a suitable base for training the detection model. As output, we get a list of relevant, man-made objects and the related cropped images.

Vessel classification into military and civilian classes is subsequently conducted through a separate model. This separation allows object detection and classification models to better focus on their respective tasks. As classification model, a modern Transformer architecture [6] is trained by adjusting the model's classification layers for the available vessel classes. The 20 most numerously represented classes in the Jane's Fighting Ship Database[4] serve as training data for this classifier. The resulting classifier generates a confidence score for each class. The relative strength of the strongest predicted class can give an initial clue towards the uncertainty of the AI model. In the depicted example, the rather civilian class training ship achieves a 28% confidence score, closely followed by 20% for the military class patrol boat. With this level of ambiguity in the model's output, no clear decision can be reached and further analysis by the human operator is merited.

Established methods for generating explanations for such classification results often rely on back-propagation approaches, such as Gradient-weighted Class Activation Mapping (Grad-CAM) [18] and its variants. However, for practical use, black box methods that do not require direct access to the internal structure of the AI model, such as Randomized Input Sampling for Explanation of Black-box Models (RISE) [16] can yield similar or even better results. The final explanations using RISE for both classes are shown in Figure 3. Using this visualization, the operator can discern that the harpoon on the whaling ship led to the suspiciously high score for the military class. Similarly, the rigging of the boat was the dominant factor contributing to its classification as civilian vessel. Using this information, the operator can identify the reason for the AI model's indecision and confidently resolve the situation.

---

[4] https://www.janes.com/capabilities/defence-equipment-intelligence/naval-combat-systems

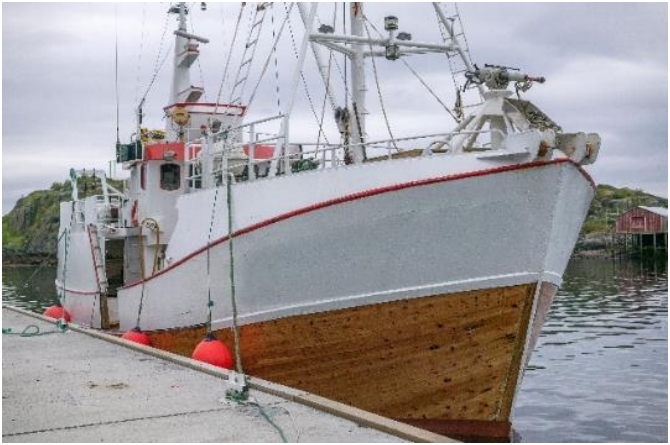

(a) Fishing boat with harpoon (© Uwe - stock.adobe.com).

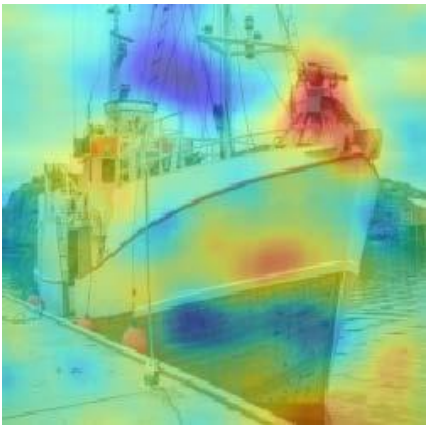 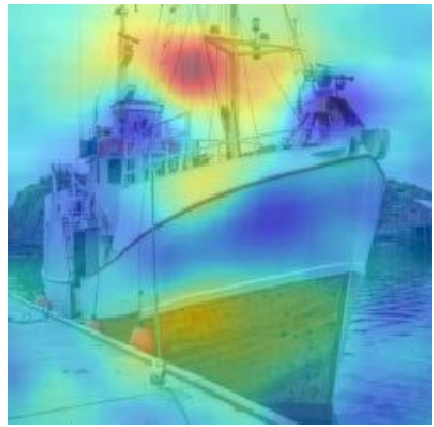

(b) Explanation for military classification.   (c) Explanation for civilian classification.

Figure 3   The fishing boat that results in an ambiguous classification result. The AI system's confidences for military or civilian classes are not strong enough to support a clear decision. The explanation heatmaps help resolve the uncertainty by pointing out the erroneous reliance on the harpoon that led to an increased confidence for the military class.

## 3.3   Applicable Ethical Principles

Considering this use case, we analyse the applicability of the ethical principles as presented in Section 2.3:

**Traceability:** In our understanding, there is a close connection between transparency and traceability. We achieve transparency by introducing XAI to each layer of the AI-based assistance system. The class confidence gives the first indication for an ambiguity that the system cannot solve without the human-in-the-loop. A visual explanation is provided by heatmaps that show the regions of interest in the image leading to the system's decision. Together with the objects' motion direction and spatial proximity, a threat analysis is conducted and intuitively visualized for the human user. By explaining and visualizing all the relevant factors that led to the AI results on each layer, the user is enabled to quickly understand the current situation and thus make an adequate decision. In addition, traceability is achieved as the system enforces a user interaction such as a confirming mouse click after each step in the process. Logging all information provided by the system and each user interaction allows a retrospective review of the scenario for further analysis and auditability.

**Proportionality:** Despite facing the immediate threat, the user is still able to question the proportionality of a decision to be taken. The design of the AI system as an assistance tool mandates that the user reflects on the recommendations before decision-making. Potential consequences are highlighted by the system in compliance with the currently valid RoE. The human-in-the-loop is still the decisive authority. However, the described assistance function shortens the reaction time to take a responsible decision and resolve the threat. In this particular case, the submarine dives deeper to avoid colliding with the civilian fishing boat instead of accidentally attacking a vessel misclassified as military. Obtaining user feedback for a crucial decision to distinguish between a civil and a military vessel guarantees the proportionality here.

**Governability**: The human-in-the-loop must be able to override an AI-supported decision at any time. The way in which the AI system is designed makes it obvious that this is the case. Assuming a negative development of the situation, the system could undoubtedly decide for the presence of a military vessel just because the corresponding confidence score is slightly higher than that for a civil vessel. If the user then could not override this decision, countermeasures could be initiated leading to a fatal misinterpretation of the actual scenario.

**Responsibility**: The human-in-the-loop is responsible for taking the final decision on how to resolve the present threat. The AI-powered assistance system provides decision support by displaying the current observation status along with its uncertainty and the possible consequences. This also shows the importance of a skilled user, who is able to visually confirm and correct the system's uncertainty. The assistance system relieves the user's workload for most of the operating time. However, the user still has to be present for handling crucial situations and edge cases. In all cases, the use of XAI methods and visualization enables the human user to understand the support provided by the AI system and, therefore, take responsibility for a decision made on the basis of the system's recommendations. Furthermore, we can identify another level of responsibility: the AI engineer enabled the system to detect and report its own uncertainty.

**Reliability**: The whaling boat seems to be unknown to the AI-based system. This can be considered as an unattended bias in the training data if we assume that the system never saw whaling boats during the learning phase. This is not surprising since whaling boats occur very rarely and only in very specific areas of the world. However, since novel or previously unseen vessel types can appear at any time (especially in the military domain), the system must be able to detect and report its own uncertainty. This is the case here since the confidence score is similarly low for multiple classes, which indicates the system's uncertainty. In addition, the system explicitly informs the user with a visual prompt about the possible consequence of a misclassification (see Figure 1). Transparency is provided by the XAI-generated heatmaps together with the highlighted confidence scores for the related classes.

# 4 Use Case 2 - Decision Support for Military Camp Protection

## 4.1 Scenario

The second scenario takes place in a military camp. The camp is surrounded by two security perimeters to protect it against intruders. An airborne Wide Area Motion Imagery (WAMI) sensor is observing the camp additionally covering several square kilometres of the surrounding environment. An officer analyses the acquired images assisted by an AI-powered system. This system contains an automatic function for object detection and trajectory analysis. The officer can also automatically align several Pan-Tilt-Zoom (PTZ) cameras to further investigate anomalies around the camp. In addition, information can by conveyed to the officer by sources such as radio communication and direct contact to other soldiers – being out-of-band for the system.

Figure 4 shows a schematic GUI of the AI-powered assistance system. The camp and its perimeters are shown on a map display. The map is grey-valued so that the colours of conspicuous trajectories within the perimeters are easier to notice. The WAMI sensor is used by the assistance system to detect and analyses moving objects within the perimeters.

On this map there are two distinct events of moving objects within the first perimeter. One single trajectory is at the south of the camp. The AI system categorizes this trajectory as a fast-driving vehicle considering the features speed and direction. An overlaying pop-up window shows the WAMI live stream for this trajectory. On the east side of the camp several trajectories are displayed. Those trajectories are categorized as a group of moving people approaching the inner perimeter. The corresponding WAMI live stream section is also shown.

The right side of the GUI shows an overview at the top for better orientation. Right below, the results of the AI-assisted threat analysis are displayed. It is based on the coarse classification of the detected objects from WAMI data in combination with a time-series analysis of their trajectories. The displayed threat results are supplemented and explained by an XAI report. The XAI window shows the feature importance and the overall threat analysis confidence of the AI system for both events. A message is shown, informing the officer that the guards got automatically notified to be on standby. To gain more insight into the threat analysis, the officer can click on the WAMI overlay images. A pop-up window with further XAI results appears as shown in Figure 5. Below the close-up WAMI image, the officer sees the prediction confidence of the AI system and the features that contributed most to this prediction. In this example, the system is sure about the categorization of moving people, but unsure about their behaviour. The AI has similar confidence for the features playing and running, just digging is lower in score but still quite possible.

The threat analysis also informs the officer that the moving people are too far away and more information is needed. The officer activates two PTZ cameras that are automatically directed towards the two events to gain more information. At the same time, the vehicle turns onto a road heading directly towards the camp. The officer is notified about this new threat characteristic by a blinking red overlay on the corresponding WAMI image. The blinking stops when the officer clicks on the WAMI image as a confirmation of noticing the new information. Similar to the first scenario, this action is logged to achieve traceability about the officer being aware of the just provided update.

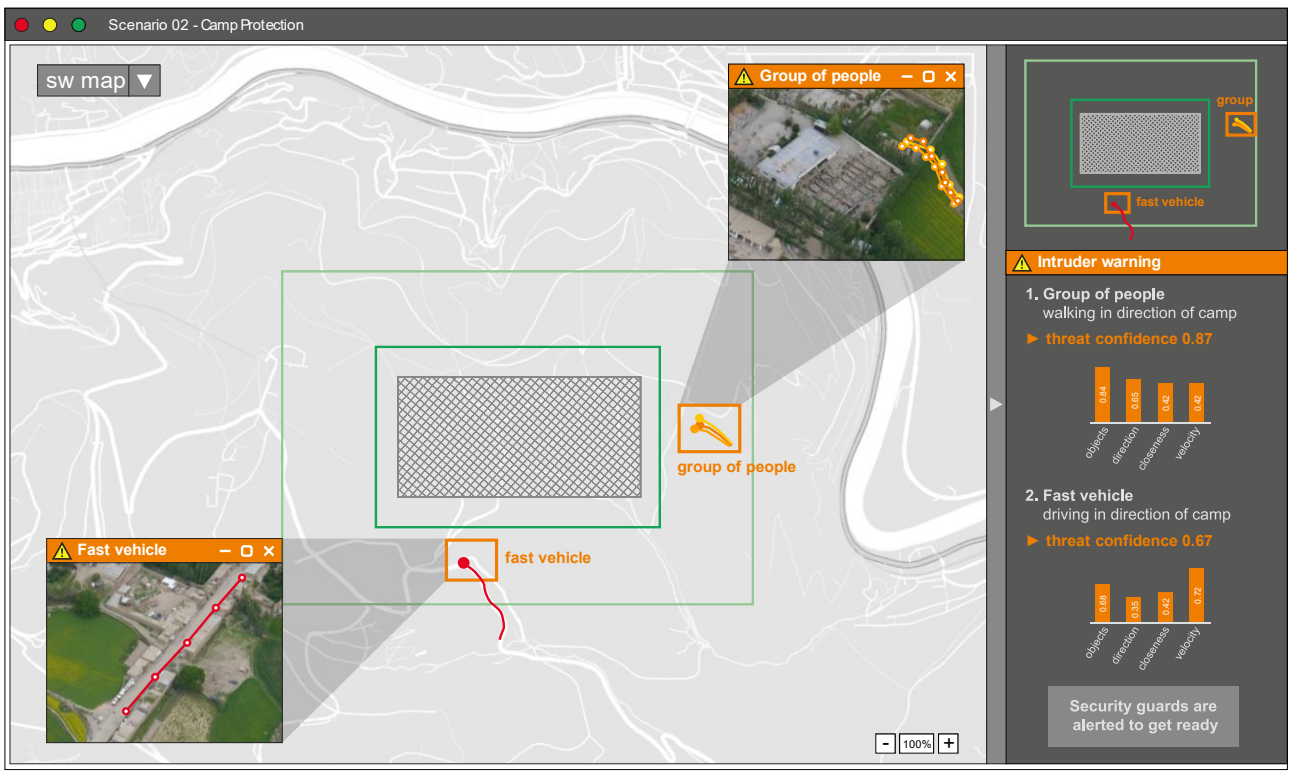

Figure 4  Schematic GUI of the AI-powered assistance system for military camp protection. The two security perimeters are depicted by a red and a black rectangle, respectively. Within the outer perimeter, two events of moving objects are detected by a WAMI sensor and coarsely classified as a fast-moving vehicle and a group of people. Pop-up windows are overlaid in the GUI for these events showing the relevant sections of the WAMI live stream. In addition, based on the WAMI data a threat analysis is performed and displayed in the right-hand side of the GUI.

Figure 6 shows the live footage of the activated PTZ cameras. With this new information, the AI-powered system and the officer are able to identify the moving people at the east of the camp as playing children. The system recommends stopping the children entering the inner perimeter by a soft intervention, i.e., informing the local police to escort the children out of the security perimeter. This recommendation is in line with the RoE for this mission. At the same time, the other PTZ camera image shows persons on a truck at the south of the camp. The system automatically recognizes that these persons are armed. Resulting from the detection of weapons in combination with the vehicle now heading directly towards the camp, a higher threat level is deduced by the threat analysis. As the threat got more severe, stronger countermeasures are now recommended. The officer again has to confirm that the information update was noticed by clicking the corresponding buttons for both events.

While a simple rule engine of the AI system makes recommendations about countermeasures in terms of the Standard Operating Procedures (SOPs) for camp protection and in accordance with the RoE of the mission, the officer is in charge to trigger and coordinate such countermeasures, like giving a warning to the camp's protection forces. The officer can also initiate countermeasures before the system makes a recommendation. This is relevant in cases, where the officer has access to additional information that the system is not aware of, such as radio communication. Conversely, the officer can decide to postpone a decision or countermeasure, in order to first acquire more information.

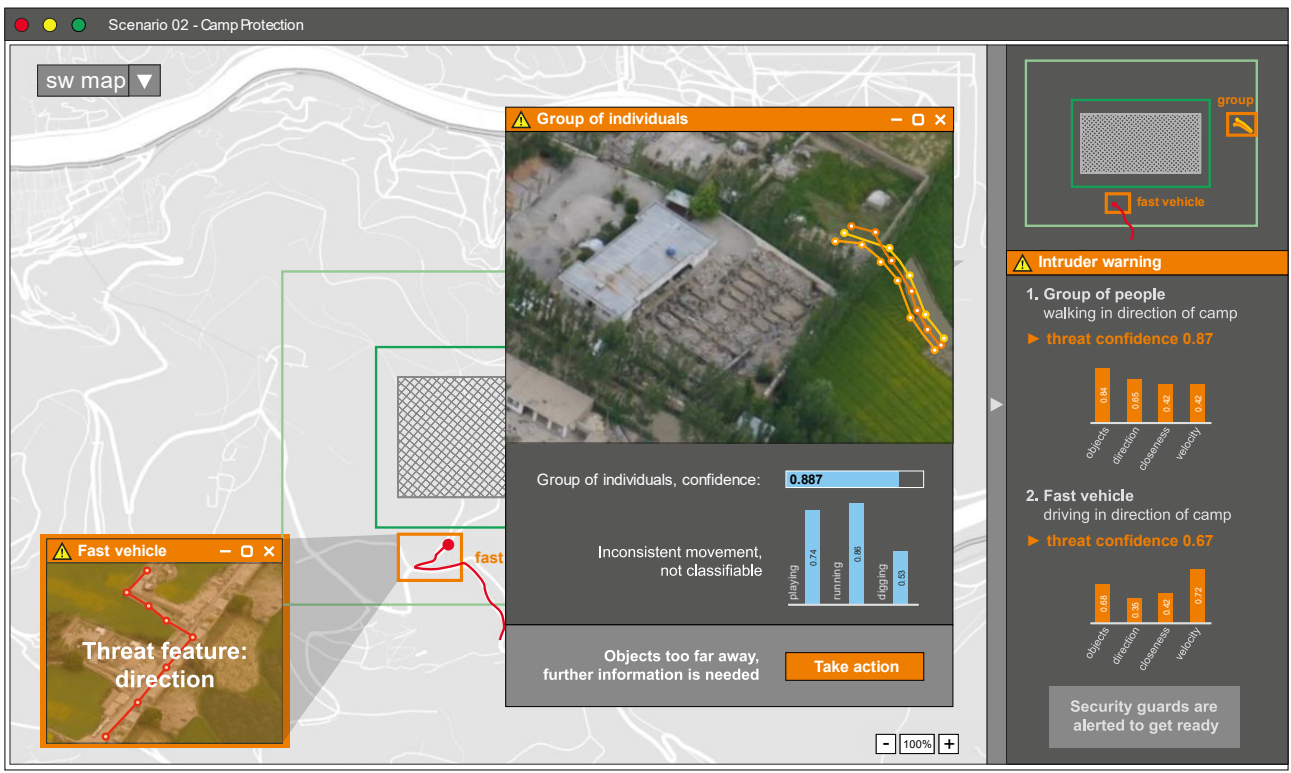

Figure 5   Detailed view of the schematic GUI for the AI-powered assistance system: Zoom-in on one threat event. To gain more insight into the displayed events, an officer can click on the overlaid images shown in Figure 4. As a result, a further pop-up is displayed, showing details of the threat analysis, including XAI results, as well as additional information on the current situation.

## 4.2   Technical Point of View

Similar to the previous maritime scenario, deep neural network models can be used for object detection in WAMI imagery as well [20]. However, objects in WAMI data are typically much smaller and employed methods thus focus on locating objects based on their motion, rather than appearance. While moving objects can be reliably located, only very coarse assumptions about object classes can be made, based primarily on the object's estimated size. A subsequent object tracking approach, such as tracking-by-detection, aggregates individual detections into trajectories. Persistent object tracking produces continuous trajectories even if vehicles stop at intersections or traffic lights [19]. The final result of this processing stage is a continuously updated set of object trajectories with rough class estimates, which can be passed on to a trajectory analysis module to determine further information.

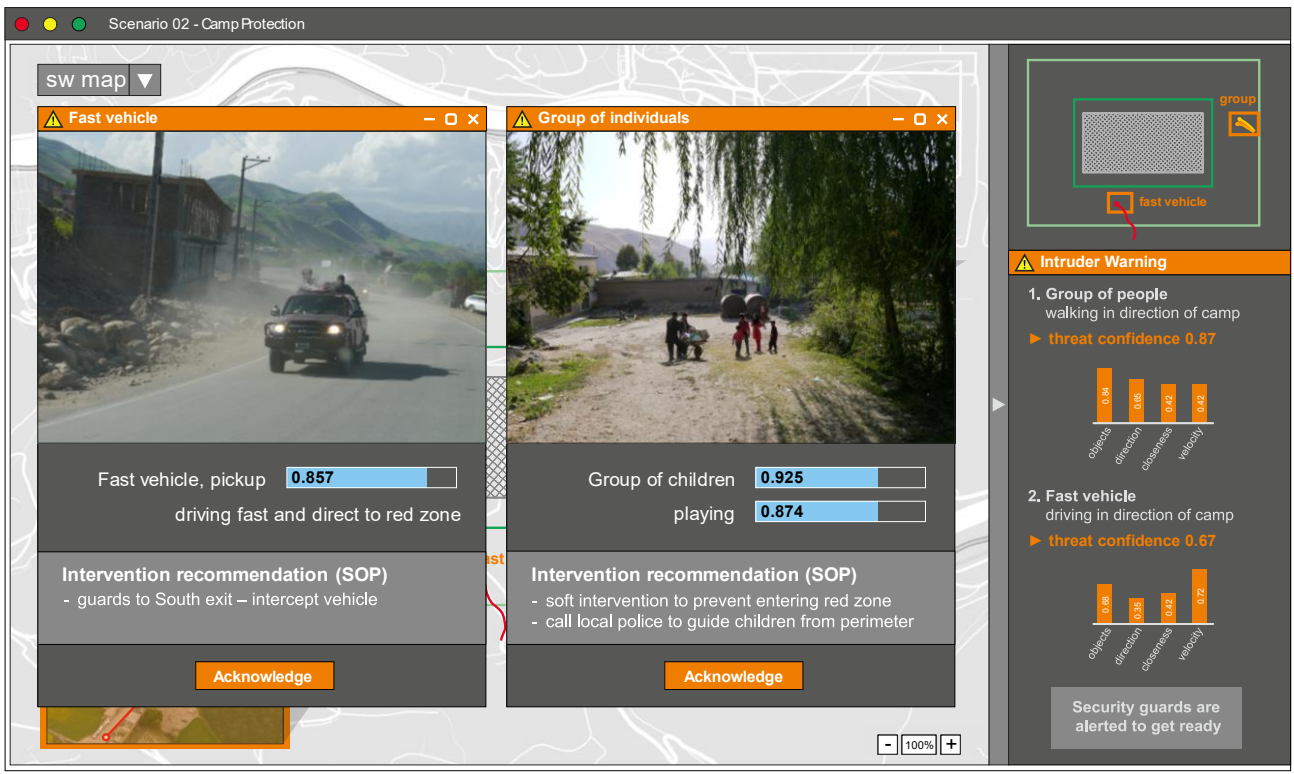

Figure 6   Live footage view (© Fabian Jeschke) of the schematic GUI for the AI-powered assistance system: for both detected events, live footage from close-ups by PTZ cameras is overlaid in the GUI. This live footage is supplemented with the results of the threat analysis and an intuitive explanation of these results as well as recommendations for adequate countermeasures.

Trajectory analysis in the context of the outlined scenario aims at identifying unusual or salient object trajectories that indicate possible threats and should be brought to the operator's attention. On the one hand, this can be implemented through the definition of restricted zones, where no object is allowed to enter. On the other hand, the pattern of life [3] can be utilized to model expected motion and behaviour of objects within the security perimeters. It considers various trajectory attributes such as direction and velocity that can serve for further analysis, including anomaly detection to identify unusual movement patterns or recognizing specific behaviours of interest [12]. The increased threat potential is then visualized in the graphical user interface to notify the officer.

For a transparent analysis of the resulting notifications, the user can rely on the automatic application of XAI methods for trajectory analysis, such as SHAP [13]. Those methods are based on a feature attribution principle meaning that the influence of the different inputs on the output is evaluated. The resulting assessment can be represented for example in the form of bar plots as seen in Figures 4, 5, and 6. Thus, these methods will result in further insight into the reasons for the trajectory being highlighted as an anomaly. For example, the result will indicate that the velocity of the object in combination with its motion direction is the primary factor for the consideration as a threat.

When the user confirms that a threat is possible, a closer look at the situation is taken through automated target handover from the WAMI sensor to a nearby PTZ camera. The higher object resolution from a side view angle provided by the PTZ cameras then allows for a fine-grained classification of object characteristics [11], such as vehicle type, e.g. pickup, or person type, e.g. children, as well as the observed activity [9], such as playing. These further details as well as the manual inspection of the PTZ imagery then enables the user to make an informed decision on how to proceed while being fully aligned with the RoE.

The recommendations for actions and countermeasures provided to the user can be derived using a rule engine. One state-of-the-art approach in this context is to employ an ontology as a knowledge model. Standards and tools as defined in the Semantic Web Stack can be used to represent the needed knowledge as a machine-readable ontology and to perform reasoning on the represented knowledge for deriving recommendations. Following this approach, the desired recommendations for countermeasures are modelled by a knowledge engineer prior to the operation of the system. In the modelling process, both the applicable SOPs and RoE have to be considered. The knowledge model relates the information acquired by the sensors and by user input to the possible recommendations.

## 4.3 Applicable Ethical Principles

Considering this use case, we analyse the applicability of the ethical principles as presented in Section 2.3:

**Traceability**: Just like in the first use case, traceability is achieved by a combination of XAI-induced transparency, enforced user interaction, and logging. The AI-powered assistance system explains its observations by highlighting salient trajectories. This saliency is produced by objects moving into the military camp's first security perimeter. Although two events of interest happen in parallel, clarity is maintained for the user by showing live footage from both events together with a clear indication of the features contributing most to the threat assessment: the number of objects, the velocity, and the motion direction. The user has to perform mouse clicks on the reports and updates to confirm that the system's notifications have been acknowledged. All reports and user interactions are logged to ensure that informed decisions were taken, and that the entire situation can be assessed and reviewed afterwards.

**Proportionality**: Decision-making by the human-in-the-loop is safeguarded at all time by the AI-powered assistance system. We can see the recommendation that more information is needed in Figure 5. As soon as a sufficient amount of information is available, possible decisions are proposed by the system in direct accordance with the currently valid SOPs and RoE. Stronger countermeasures are recommended only in the case when a threat becomes more direct or immanent such as the pickup truck that suddenly approaches the camp with a high velocity.

**Governability**: Decisions can be postponed if more information is needed. This is not only the case if the AI-powered assistance system recommends collecting more information, but also if the user does so. On the other hand, the user can take actions even if the system suggests waiting and thus it can be overruled at any time. As a result, the human-in-the-loop can bypass the system and make an instant decision. This can be relevant if the user gained situational awareness by additional sources of information that the AI-powered assistance system does not have such as direct radio communication to a human scout in the field.

**Responsibility**. The user can react to the notification of a possible threat by contacting the camp security at the user's discretion. The alerting by the system becomes more urgent to signalize that the reaction time gets shorter and shorter. However, the human-in-the-loop is still the final and sole decision-maker.

**Reliability**: The AI-powered assistance system is able to detect multiple possible threats emerging in parallel. It also understands the need to gain more information before a decision can be recommended. The system employs XAI methods to justify to the user how the recommendation was generated, thereby enabling the user to verify its plausibility. Hence, the system's uncertainty is transparently communicated to the user and actions to resolve the uncertainty are suggested. The threat level develops over time and thus gives the user the chance to understand the context of the possible threat. The alignment with the currently valid SOPs and RoE guarantees informed decisions.

## 5 Use Case 3 - Land-based Reconnaissance in Inhabited Area

### 5.1 Scenario

The third scenario takes place in an urban area. The urban area is divided by a river, which serves as a border between a demilitarized zone on one side, and own territories on the other side. The area is characterized by a large number of moving objects, making it crucial to thoroughly assess the threat level for the objects to effectively mitigate potential harm to civilians and critical infrastructure. Therefore, a reconnaissance vehicle is positioned equipped with advanced surveillance technology, such as high-resolution cameras, mobile sensors, and an AI-based assistance systems. The vehicle is located at an elevated position with a wide view of the entire area to detect potential threats and provide relevant information to the command.

Figure 7 shows a schematic overview of the scenario and the events. The urban area and its perimeters are shown on a map display in the first frame. A wide area surveillance camera system is used by the assistance system to detect and analyse moving objects within the perimeter. An AI assistant informs the operator of relevant events. This AI assistant is accessible via a chat-like window at the right, overlaying the displayed map and observations. In the first frame the AI assistant system detects an enemy armoured vehicle, and an immediate notification is sent to the operator. To gain the operator's attention, standard messenger notification approaches are applied. In the chat window, the AI assistant states the observation in simple natural language for fast and intuitive understanding. At the same time, the detected vehicle is overlayed on the map. In addition, the AI assistant offers suggestions the analysis of the situation. Those suggestions can help the operator to assess the seriousness of the threat and identify the best course of action. The AI system serves as a valuable tool in navigating the complex landscape of regulations and guidelines such as RoE and SOPs that govern the safety and well-being of civilians in such dynamic environments. The system tracks the movement of the vehicle in the background, displaying its trajectory on the map. By continuously monitoring the vehicle's movements, the system provides real-time updates on its location and trajectory. In the second frame of Figure 7, the assistant informs the user that the vehicle has left the demilitarized zone and has come to a stop near a school. In response, the user wants to know if the vehicle is now considered an eligible target. The assistant, referring to the relevant SOPs, confirms that it is not due to the school building nearby.

Simultaneously, a second AI assistant instance contacts the operator through a chat interface indicated by the red messenger icon. This second assistant provides the user with important information regarding another observation of a vehicle as shown in third and fourth frame of Figure 7. The second assistant conducts a situation analysis and predicts an increasing threat level. In response to the situation and the heightened threat prediction, the operator seeks further clarification from the AI assistant regarding the available course of action. By analysing the vehicle's trajectory, the assistant predicts a specific area that is likely to be crossed and that meets the criteria for indicating the vehicle as valid target without harming protected infrastructure or civilians according to the RoE and SOPs. This simultaneous activity of multiple AI assistants ensures comprehensive monitoring and analysis of the evolving scenario. By alerting the operator about critical information and visualizing the threat level, these AI assistants play a significant role in supporting the operator making informed decisions and taking appropriate actions.

Meanwhile, the fifth frame of Figure 7 shows an update on the first observation. In this conversation, the AI assistant informs the user about the detection of movement around the enemy armoured vehicle. However, due to the large distance between observer and scene and the limited sensor resolution, the AI assistant is unable to provide a more precise analysis of the observed movement. Recognizing the need for additional information and a closer examination of the situation, the AI assistant recommends sending an autonomous tactical unmanned aerial system (UAS). In this way, sensor resource management is performed to fill information gaps efficiently with relevant data acquired by available sensor platforms. While the UAS is gathering data, the second AI assistant reports to the operator that the hostile convoy has surpassed the designated coordinates and is now considered a valid target. The user verifies this and approves the target.

This situation is now resolved and the chat icon turns grey to indicate this. The operator can archive the chat for a cleaner GUI. The archived data can be used for retrospection. Switching back to the vehicle in front of the school: the UAS is equipped with AI image analysis capabilities, enabling it to perform a motion analysis of the captured footage. Through the AI assistant, the UAS reports the detection of wounded soldiers back to the operator. Again, the AI assistant acquires relevant RoE and SOPs. Based on their analysis, the assistant informs the operator that there is no valid target present and assures that the safety of children or wounded humans must not be endangered.

## 5.2 Technical Point of View

In this use case the decision support features are (1) the core of the AI assistant, (2) aiding users in considering relevant RoE and SOPs, and (3) rules during stressful situations. There are several approaches to realize such an AI-assistant. Symbolic AI techniques such as knowledge models and decision trees provide a solid foundation for embedding expert knowledge in AI-systems. Additionally, these approaches also offer the possibility of intrinsic interpretability. Knowledge models such as knowledge graphs serve as instruments that enable AI systems to manage complex behavioural rules such as RoE and SOPs. In this way, AI can help users by highlighting the applicable rules for given situations and indicating the allowed actions.

For an effective implementation, a detailed modelling of RoE and SOPs is essential. This involves creating a comprehensive representation of the rules that govern the conduct of operations and engagements. To align real-world situations with the appropriate rules, AI-based systems must be capable of instantiating elements from the knowledge model based on processed sensor data. Consequently, the knowledge model must incorporate a comprehensive catalogue of objects and structured that sensors are likely to observe as well as their relation to these rules. Such a catalogue could contain a detailed hierarchy of military vehicles, different roles of humans such as military, civilian, wounded, or adolescent persons, and meta data of buildings and critical infrastructure such schools, hospitals, power plants, or bridges. Employing this knowledge model together with sensor data analysis, the AI system is tasked with assisting users by reporting relevant observations, highlighting applicable rules, and suggesting potential actions and decisions. This methodology ensures that AI assistance is both contextually relevant and adherent to established protocols.

However, extensive knowledge models are difficult to implement and generalize. Modern sub-symbolic AI techniques, particularly Large Language Model (LLM), offer a significant advantage in terms of generalizability, yet they lack interpretability. Building upon the knowledge model framework, Retrieval Augmented Generation (RAG) can further enhance the AI system's ability to deliver pertinent rules for recognized real-world situations. RAG enhances the response generation capabilities of language models by incorporating external information [8]. It works by first retrieving relevant documents or data from a great source of information such as a database or the internet. Then, this information is used to inform the generation process of the language model. This allows the model to produce more accurate, informative, and contextually relevant outputs, as it can draw upon a broader range of knowledge than it has been explicitly trained on. RAG is particularly useful in applications where up-to-date or domain-specific information is required to answer questions or solve problems.

In the context of the previously described system, RAG can complement knowledge models by pulling in additional, contextually relevant information that may not be explicitly encoded in the knowledge graph. This is particularly advantageous in rapidly evolving scenarios where the pre-defined rules in the knowledge model may not be sufficient. However, RAG's reliance on external data sources can also introduce challenges. The quality of the AI's output is contingent on the quality and relevance of the retrieved information. If the external data is outdated, biased, or incorrect, it could lead to inappropriate suggestions. Moreover, RAG systems require sophisticated filtering mechanisms to ensure that the retrieved information is indeed applicable to the current situation, avoiding the risk of information overload or irrelevant data complicating decision-making processes.

## 5.3 Applicable Ethical Principles

Considering this third use case, we also examine the applicability of our ethical principles in Section 2.3:

**Traceability:** The AI assistant provides detailed reports on its observations such as identifying a hostile armoured vehicle near a school. These reports include visual overlays on the map display, live footage, and explanations of threat assessments. Additionally, all operator interactions, including decisions made based on the AI's recommendations, are logged for later review. By ensuring that all actions and decisions are transparently documented, the system enables stakeholders to trace the reasoning behind each decision, fostering accountability and trust in the AI's capabilities.

**Proportionality:** When the AI assistant detects a potential threat such as a hostile vehicle approaching a civilian area, it recommends responses that are proportional to the threat level while minimizing harm to civilians. This is done in accordance with the SOPs and RoE. For instance, it may advise against immediate engagement if there is a risk of collateral damage, opting instead for surveillance or non-lethal measures. By advising proportional responses to threats, the system helps maintaining a balance between achieving mission objectives and minimizing risks to civilian lives and infrastructure, thereby upholding ethical standards in urban reconnaissance operations.

**Governability:** The AI assistant allows operators to exercise control over the decision-making process by providing options for postponing actions when additional information is needed. Operators can also override the AI's recommendations based on their own judgment or situational awareness, ensuring that human control remains paramount. How it relates: by facilitating human oversight and intervention, the system ensures that decisions are made in accordance with operational guidelines and ethical considerations, enhancing the effectiveness and accountability of urban reconnaissance operations.

**Responsibility:** Although the AI assistant offers recommendations and insights, the ultimate responsibility for decision-making remains with the human operator. Operators must carefully consider the AI's suggestions in light of operational guidelines, ethical principles, and the safety of civilians and infrastructure before making final decisions. By emphasizing human responsibility in decision-making, the system promotes accountability and ethical conduct in urban reconnaissance operations, ensuring that actions align with mission objectives and respect fundamental ethical principles.

**Reliability:** The AI-powered system demonstrates reliability by detecting multiple potential threats, understanding the need for more information before making recommendations, and employing XAI methods. The transparency of the system's uncertainty, continuous monitoring of threat levels, and alignment with the SOPs and RoE contribute to informed and reliable decision-making by the user.

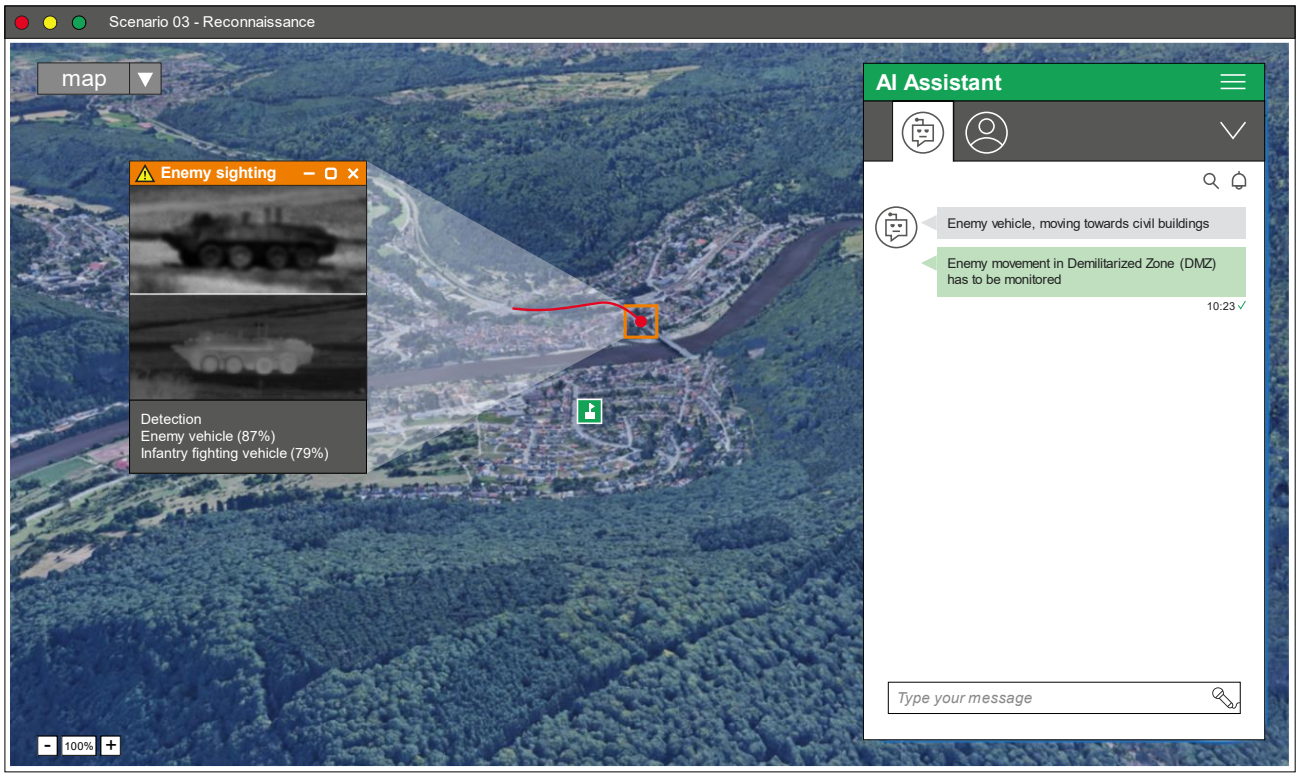
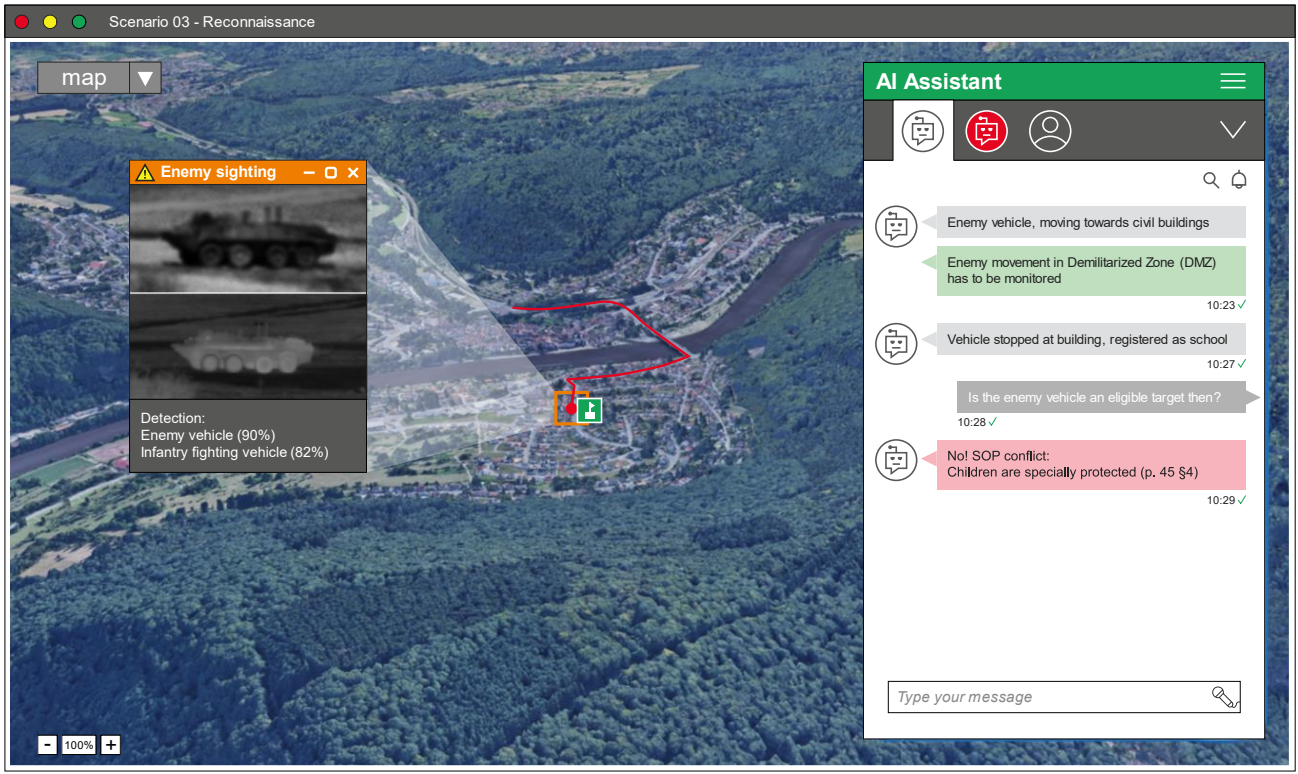

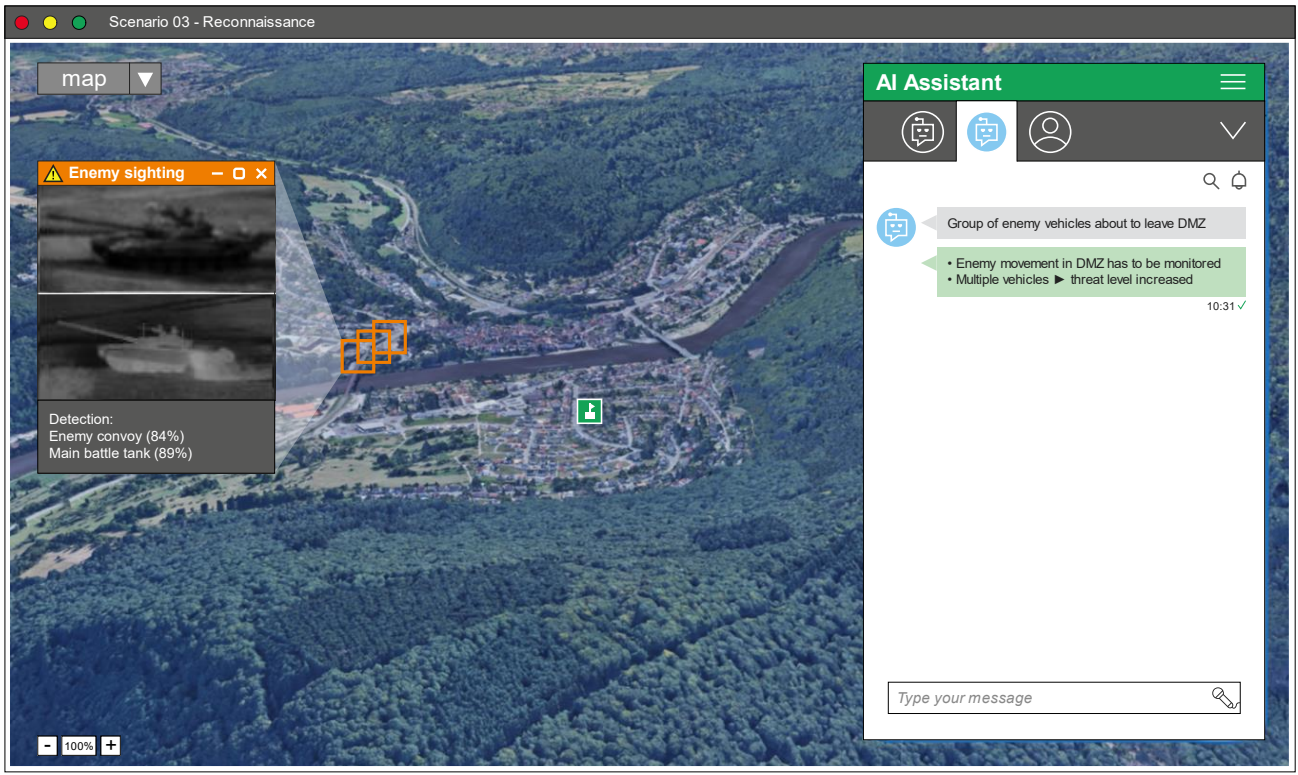
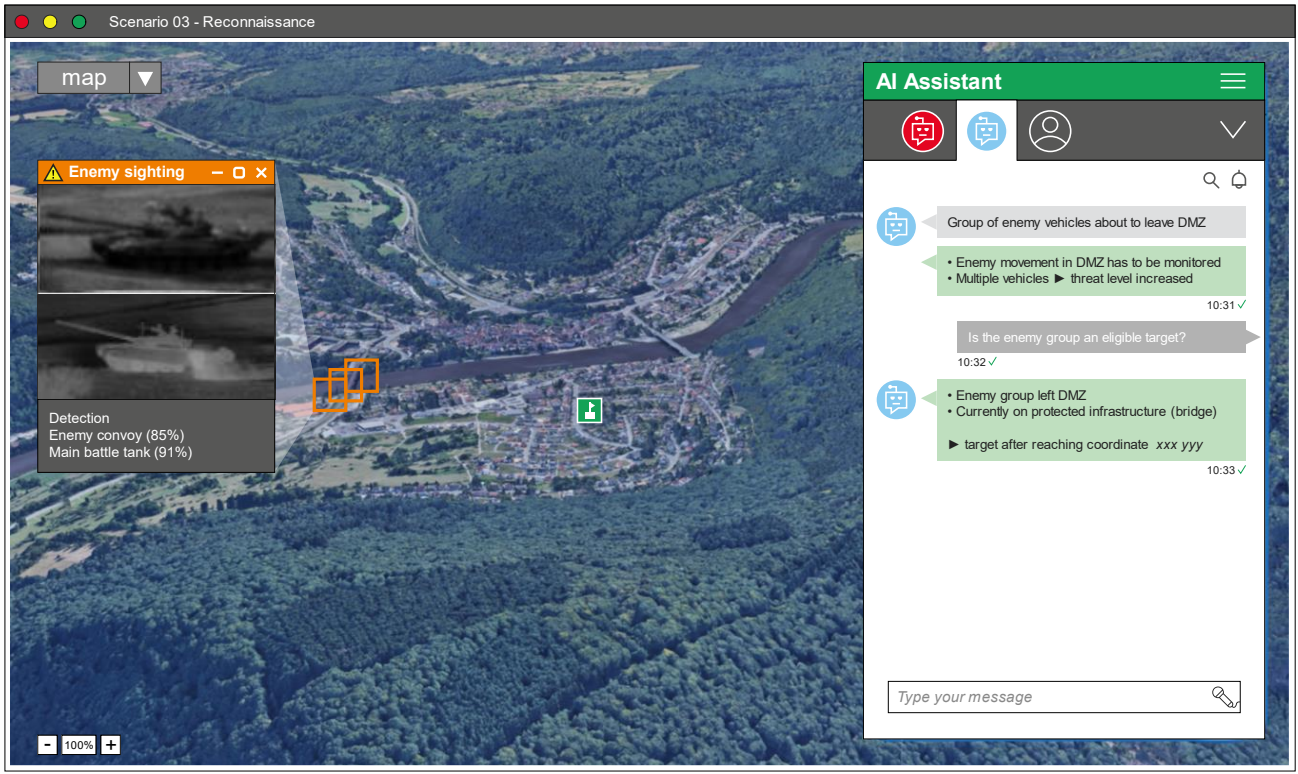

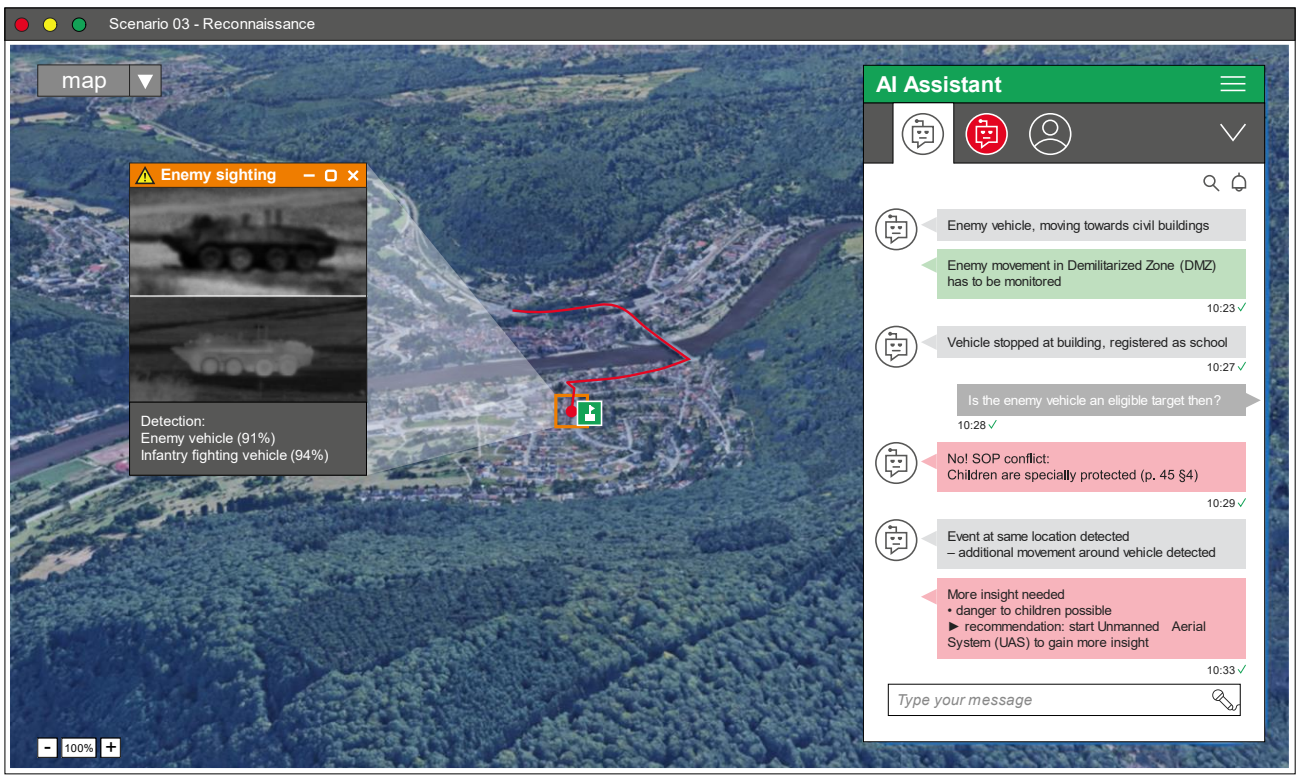
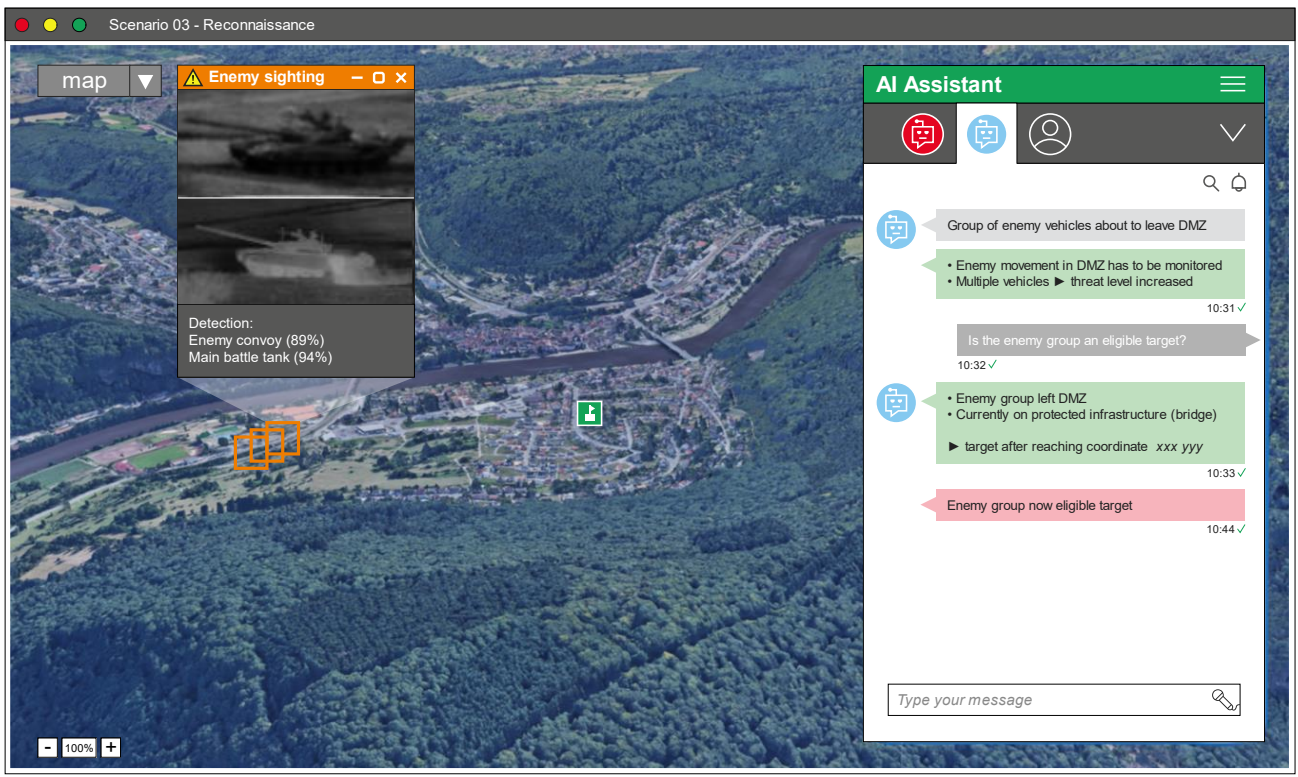

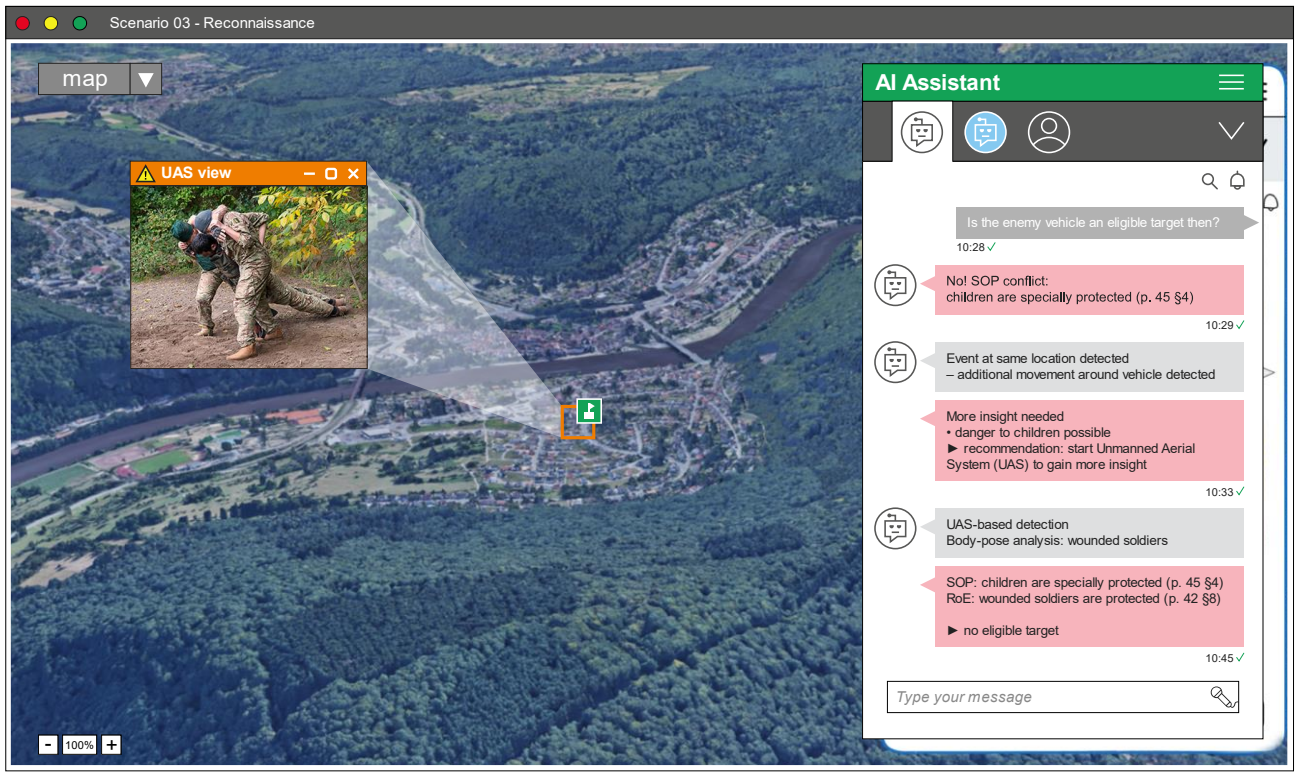

Figure 7   Schematic of the third scenario: an AI-powered assistance system supports the operator in a chat-like manner. In this way, an eligible target can be identified while implicitly considering Rules of Engagement (RoE) and Standard Operating Procedures (SOP). Pictures taken from [25] and from Adobe Stock (© Marina - stock.adobe.com).

# 6 Conclusion

This white paper underscores the critical importance of responsibly deploying AI in military contexts, emphasizing a commitment to ethical and legal standards. The evolving role of AI in the military goes beyond mere technical applications, necessitating a framework grounded in ethical principles. The discussion within the paper delves into ethical AI principles, particularly focusing on the Fairness, Accountability, Transparency, and Ethics (FATE) guidelines. Noteworthy considerations encompass transparency, justice, non-maleficence, and responsibility. Importantly, the paper extends its examination to military-specific ethical considerations, drawing insights from the Just War theory and principles established by prominent entities such as the U.S. Department of Defence (DoD), NATO, and the U.K. Defence Science and Technology Lab (DSTL).

In addition to the identified principles, the paper introduces further ethical considerations specifically tailored for military AI applications. These include traceability, proportionality, governability, responsibility, and reliability. Traceability is emphasized to ensure transparency, allowing a clear understanding of the decision-making processes within AI systems. Proportionality aligns AI outcomes with both humanitarian and military objectives, emphasizing a balance in the application of AI capabilities. Governability underscores the need for human control over AI systems, emphasizing the ability for human operators to override decisions. Responsibility places accountability squarely on human decision-makers, acknowledging their ultimate role in decision-making. Lastly, reliability addresses the robustness of AI methods, recognizing the importance of ensuring dependable and trustworthy AI systems for military use. The application of these ethical principles is discussed on the basis of three use cases in the domains of sea, air, and land. Methods of automated sensor data analysis, XAI, and intuitive user experience are utilized to specify the use cases close to real-world scenarios.

This comprehensive approach to ethical considerations in military AI reflects a commitment to aligning technological advancements with established ethical frameworks. It recognizes the need for a balance between leveraging AI's potential benefits in military operations while upholding moral and legal standards. The inclusion of these ethical principles serves as a foundation for responsible and accountable use of AI in the complex and dynamic landscape of military scenarios.